\documentclass[aps,prl,twocolumn,superscriptaddress,showpacs,floatfix]{revtex4}
\usepackage{graphicx,epsfig,amsmath}

\begin{document}

\title{The Effect of Electrostatic Screening on a Nanometer Scale Electrometer}

\author{K. MacLean}
    \email{kmaclean@mit.edu}
    \affiliation{Department of Physics, Massachusetts Institute of Technology, Cambridge, Massachusetts 02139}
\author{T. S. Mentzel}  
    \affiliation{Department of Physics, Massachusetts Institute of Technology, Cambridge, Massachusetts 02139}
\author{M. A. Kastner}
    \affiliation{Department of Physics, Massachusetts Institute of Technology, Cambridge, Massachusetts 02139}


\begin{abstract}
We investigate the effect of electrostatic screening on a nanoscale silicon MOSFET electrometer.  We find that
screening by the lightly doped $p$-type substrate, on which the MOSFET is fabricated, significantly affects the 
sensitivity of the device.  We are able to tune the rate and magnitude of the screening effect by varying the temperature and the voltages applied to the device, respectively.  We show that despite this screening effect, the electrometer is still very sensitive to its electrostatic environment, even at room temperature.
\end{abstract}

\pacs{73.63.-b, 85.35.-p, 85.30.-z}

\maketitle

Nanoscale electrometers have emerged as powerful tools for studying a wide variety of solid state systems.
These sensors can be integrated on a semiconductor chip adjacent to a solid state structure of interest \cite{Field1993:NoninvasiveProbe}, or mounted on a scanning probe tip \cite{YacobyImage}.  Utilized in these configurations, nanoscale electrometers have had a great impact on the study of single electron devices \cite{Elzerman2004:SingleShotReadOut,Petta2005:CoherentManipulation, Fujisawa:Bidirectional,Ensslin:CountingStat,SamiT1}, disordered materials \cite{Drndic:EFM,MacLeanTSF}, and high mobility two dimensional electron gases \cite{YacobyFCharge,YacobyGraphene}.  The small size of these electrometers can lead to high charge sensitivities \cite{SchoelkopfChargeSens}, which are central to many of these applications.  It is widely recognized that, of the many factors that may limit the sensitivity of a nanoscale electrometer, electrostatic screening is likely to be one of the most important.  However, because in most cases the effect of the screening is more or less fixed, and cannot be easily tuned, there have been few if any experimental investigations of this effect.

\begin{figure}
\setlength{\unitlength}{1cm}
\begin{center}
\includegraphics[width=8.0cm, keepaspectratio=true]{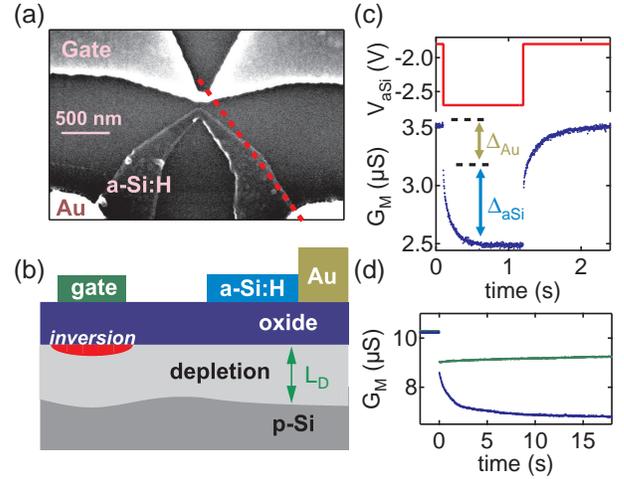}
\end{center}

\caption{(a) Electron micrograph of MOSFET gate , a-Si:H strip, and gold contacts. (b) Sketch of the cross-section of the device along the dashed red line in (a).  When a positive voltage is applied to the gate, an inversion layer forms at the Si-SiO$_{2}$ interface. A depletion region forms in the $p$-type silicon substrate beneath the Si-SiO$_{2}$ interface, as discussed in the main text. The depth of the depletion region below the Si-SiO$_{2}$ interface is denoted $L_{D}$.  (c) Voltage sequence applied to one of the gold contacts (top trace) and the conductance of the MOSFET in response to changes in charge on the gold ($\Delta_{Au}$) and a-Si:H ($\Delta_{aSi}$) (bottom trace), at T = 125 K, as discussed in the main text. (d) Result of stepping the voltage applied to the gold contacts at T = 79 K for a device in which the strip of a-Si:H is connected to only one of the two gold contacts, as discussed in the main text.  For the blue (green) data the gold contact connected (not connected) to the a-Si:H strip is changed. For these data $V_{sub}$ = 0 V.}
\label{fig:fig1}
\end{figure}

In this Letter, we characterize the effect of electrostatic screening on the sensitivity of a nanoscale MOSFET (metal-oxide-silicon field-effect-transistor) electrometer.  For our device, we find that screening by the lightly doped $p$-type silicon substrate, on which the MOSFET is fabricated, significantly affects the charge sensitivity of the device.  However, because this screening is caused by a lightly doped semiconductor as opposed to a metal, we are able to tune both the rate and the magnitude of the screening effect \textit{in situ} by varying the temperature and depth of the depletion region in the substrate, respectively.  This tunability allows us to quantify the effect of screening for our system.  We demonstrate that, despite the effects of electrostatic screening, our nanoscale electrometer can still detect very small charge fluctuations, even at room temperature.

The device used in these experiments has been discussed previously \cite{MacLeanTSF}, and consists of a nanometer scale silicon MOSFET that is electrostatically coupled to a strip of hydrogenated amorphous silicon (a-Si:H).  An electron micrograph of the device is shown in Fig. \ref{fig:fig1}(a). The $n$-channel MOSFET is fabricated using standard CMOS techniques on a silicon substrate. The substrate is lightly doped $p$-type with boron ($N_{B} \approx$ 3 $\times$ 10$^{15}$ cm$^{-3}$). Adjacent to the gate of the MOSFET, we nanopattern a strip of phosphorous doped a-Si:H. We make electrical contact to the a-Si:H using two gold contacts, which are visible as the bright regions in the two lower corners of the electron micrograph in Fig. \ref{fig:fig1}(a). For all of the work discussed here, a positive voltage is applied to the gate of the MOSFET, so that an inversion layer forms at the Si-SiO$_{2}$ interface beneath the gate, as shown in Fig. \ref{fig:fig1}(b).  The conductance of the MOSFET inversion layer, $G_{M}$, is limited by its narrowest portion, which is located underneath the $\approx$ 60 nm wide constriction in the gate.  Electrical contact is made to the inversion layer through two degenerately doped $n$-type silicon regions located on either side of the constriction (not shown in the micrograph).  We measure $G_{M}$ by applying a small voltage $\sim$ 5 mV to one contact, and measuring the current that flows out through the other.  We make electrical contact to the $p$-type substrate through the back of the chip. For the data reported below, we negatively bias the $p$-type substrate by $V_{sub}$ = -3 V relative to the $n$-type contacts unless otherwise indicated.

The conductance of the MOSFET is extremely sensitive to its electrostatic environment. In particular, $G_{M}$ is sensitive to changes in charge in either the a-Si:H or the gold contacts.  As we show below, this sensitivity is significantly affected by screening by the $p$-type silicon substrate: If charge $Q$ is added to the a-Si:H or gold contacts, an oppositely charged region will form in the substrate underneath, thereby reducing the effect of $Q$ on $G_{M}$.  This screening charge is located at the Si-SiO$_{2}$ interface, or, if the silicon beneath the
Si-SiO$_{2}$ interface is depleted of holes (Fig. \ref{fig:fig1}(b)), the screening charge will be located a distance $L_{D}$ beneath the Si-SiO$_{2}$ interface.

Our measurement consists of stepping the voltage $V_{aSi}$ applied to one of the a-Si:H gold contacts while simultaneously monitoring $G_{M}$.  An example is shown in Fig. \ref{fig:fig1}(c). Here we set the voltage applied to one gold contact to 0 V, and apply the voltage sequence shown in the top trace of Fig. \ref{fig:fig1}(c) to the other contact \cite{substrate}.  The bottom trace of Fig. \ref{fig:fig1}(c) shows the variation in $G_{M}$ in response to the voltage sequence. When $V_{aSi}$ is first stepped from -1.8 V to -2.7 V, $G_{M}$ drops by an amount $\Delta_{Au}$ in a time too short to measure, and then decreases slowly by an amount $\Delta_{aSi}$.  

As we have demonstrated in MacLean \textit{et al.} \cite{MacLeanTSF}, the slow change $\Delta_{aSi}$ in  $G_{M}$ is caused by the slow addition of negative charge to the a-Si:H.  The MOSFET electrometer senses this change in charge electrostatically, and $G_{M}$ decreases as negative charge is added to the a-Si:H.  The time scale of this charging is a direct measurement of the resistance of the a-Si:H strip \cite{MacLeanTSF}.  The much more rapid drop $\Delta_{Au}$  in $G_{M}$ is caused by the negative charge added to the gold contacts, which charge up very quickly because of their low electrical resistance. When $V_{aSi}$ is returned to -1.8 V, the same responses $\Delta_{Au}$ and $\Delta_{aSi}$ are observed but with the opposite sign, as negative charge is now removed from the gold and the a-Si:H.  A similar response is observed when the voltage sequence is applied to the other gold contact, or to both contacts at the same time.

\begin{figure}
\begin{center}  
\includegraphics[width=8.0cm, keepaspectratio=true]{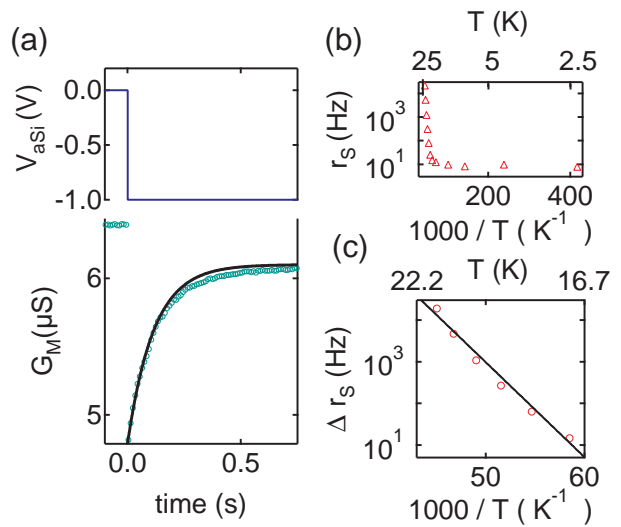}
\end{center}
    \caption{(a) Observation of the screening effect at T = 9.8 K, as discussed in the main text.  The top trace shows the voltage step applied to the a-Si:H gold contact.  For the lower trace, the solid black curve is a fit to an exponential, as discussed in the main text. (b) Screening rate $r_{S}$ as a function of inverse temperature.  (c) Change in screening rate $\Delta r_{S}$ as a function of inverse temperature, as described in the main text.  The solid line is a theoretical fit described in the main text. For all of these data, $V_{sub}$ = 0.}
    \label{fig:fig2}
\end{figure}

To confirm that our interpretation of the data is correct, we study a separate device where, like the device shown in Fig. \ref{fig:fig1}(a), a strip of a-Si:H is patterned adjacent to a nanoscale MOSFET.  However, for this device, the strip of a-Si:H is connected to only one of the two gold contacts.  The data is shown in Fig. \ref{fig:fig1}(d).  At $t = 0$ we step one contact from 0 to -9.9 V, while the other contact is held constant at 0 V.  A rapid drop $\Delta_{Au}$ is observed when the pulse is applied to either one of the gold contacts, but the slower response $\Delta_{aSi}$ is only observed when the pulse is applied to the gold which is connected to the strip of a-Si:H, confirming our interpretation of the data. 

The sensitivity of $G_{M}$ to its electrostatic environment depends on screening by the underlying $p$-type silicon substrate.  To demonstrate this, we examine the response of the MOSFET to changes in charge in the gold contacts at a temperature $T \approx$ 10 K, lower than the temperature at which the data shown in Fig. \ref{fig:fig1} are acquired.  At this temperature, the a-Si:H is so resistive that it does not charge up on the time scale of the experiment \cite{MacLeanTSF}, so that we can add charge to the a-Si:H gold contacts but not to the a-Si:H itself.  The results are shown in Fig. \ref{fig:fig2}(a). When we change the voltage applied to the a-Si:H gold contacts from 0 to -1 V (top trace), we see a large decrease in the MOSFET conductance, which gradually dies away as time progresses (bottom trace). 

%

The gradual dying away of the decrease in $G_{M}$ can be understood in terms of screening.  When we add charge to the gold contact, an opposing charge in the $p$-type substrate is induced, reducing the overall effect on $G_{M}$.  At low temperatures, the resistance of the substrate is high, and this charge is induced at a slow rate.  To quantify this rate, we fit the $G_{M}$ trace to an exponential $G_{M}(t) = G_{\infty} + G_{scr}e^{-r_{S}t}$, where $G_{\infty}$ and $G_{scr}$ are constants that depend on the voltages applied to the MOSFET gate, $p$-type substrate, and gold electrodes, and $r_{S}$ is the screening rate.

To show that this screening effect is caused by the $p$-type silicon substrate, we measure $r_{S}$ as a function of temperature.  The results are shown in Fig. \ref{fig:fig2}(b).  As the temperature is reduced, $r_{S}$ drops, saturating at a minimum value $r_{min} \approx$ 8 Hz.  In Fig. \ref{fig:fig2}(c), we plot $\Delta r_{S} = r_{S}$ - $r_{min}$ as a function of inverse temperature, and fit to an activated temperature dependence $\Delta r_{S} \propto e^{-E_{A}/kT}$.  We obtain $E_{A}$ = 45 $\pm$ 5 meV, which agrees well with the boron acceptor binding energy \cite{Sze}.  For boron-doped silicon with no donor compensation, the Fermi level lies between the valence band and the boron donor level, and the activation energy for hole transport is therefore half of the boron acceptor binding energy.  However, at sufficiently low temperatures, a small concentration of compensating donor states caused by defects or impurities $N_{D}$ will move the Fermi level into the acceptor band \cite{Shklovskii}.  In our case, the number of defects required is only $N_{D} \sim $ 10$^{10}$ cm$^{-3}$.  Because the required density is so small, we expect the Fermi level to lie in the acceptor band, and the activation energy required for the generation of holes in the valence band to be the boron acceptor binding energy.  The correspondence between the activation energy for the screening and the boron acceptor binding energy demonstrates that the conductivity of the boron doped substrate limits $r_{S}$.  Presumably $r_{S}$ saturates at a minimum value $r_{min}$ because some conduction mechanism other than activation of holes in the $p$-type substrate dominates at low temperature.  It is possible that this low temperature conduction occurs via tunneling of electrons between acceptor states \cite{Shklovskii} in the $p$-type substrate. In any case, from this data it is clear that screening by holes in the boron doped substrate significantly reduces the sensitivity of the MOSFET.  

\begin{figure}
\setlength{\unitlength}{1cm}
\begin{center}
\includegraphics[width=7.0cm, keepaspectratio=true]{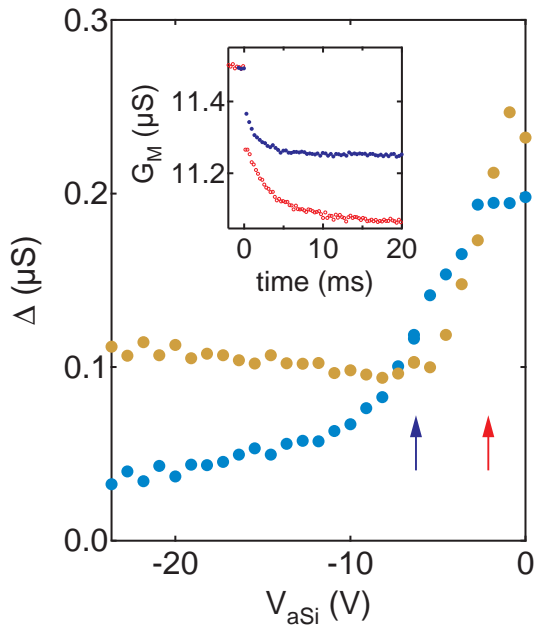}
\end{center}
\caption{$\Delta_{aSi}$ (blue circles) and $\Delta_{Au}$ (gold circles) measured as a function of $V_{aSi}$ at T = 139 K, as discussed in the main text. For these data, we make the MOSFET gate voltage more positive as $V_{aSi}$ is made more negative so that $G_{M} \approx 11$ $\mu$S at the start of each $G_{M}(t)$ trace. (Inset) Examples of data from which $\Delta_{aSi}$ and $\Delta_{Au}$ are extracted for two different $V_{aSi}$ values.  For both $G_{M}(t)$ traces, $V_{aSi}$ is stepped by -0.5 V at $t = 0$. The data are offset vertically by a small amount for clarity. The blue and red data sets are taken at the positions of the blue and red arrows, respectively.  The decrease in both $\Delta_{aSi}$ and $\Delta_{Au}$ with increasingly negative $V_{aSi}$ is clearly visible.}
    \label{fig:fig3}
\end{figure}

At higher temperatures $T > 25$ K, $r_{S}$ becomes too fast for us to measure.
In this regime, we investigate the dependence of $\Delta_{aSi}$ and $\Delta_{Au}$ on $V_{aSi}$.  The results are shown in Fig. \ref{fig:fig3}.  Here we step the voltage applied to both gold contacts from $V_{aSi}$ to $V_{aSi} - \Delta V$, where
$\Delta V$ = 0.5 V.  We extract $\Delta_{aSi}$ and $\Delta_{Au}$ from the resulting $G_{M}(t)$ trace as depicted in 
Fig. \ref{fig:fig1}(c).  We measure both $\Delta_{aSi}$ and $\Delta_{Au}$ as a function of $V_{aSi}$ and find that both of these quantities decrease as $V_{aSi}$ is made more negative. The decreases in $\Delta_{aSi}$ and $\Delta_{Au}$ are clearly visible when the $G_{M}(t)$ traces taken at different $V_{aSi}$ values are compared, as is shown in the inset to Fig. \ref{fig:fig3}.

 These results can be understood in terms of screening by the $p$-type substrate in the following way:  At $V_{aSi}$  = 0 V, the $p$-type substrate beneath the Si-SiO$_{2}$ is depleted, as depicted in Fig. \ref{fig:fig1}(a).  As $V_{aSi}$ is made more negative, $L_{D}$ is reduced beneath the gold and the a-Si:H.  This has the effect of making the screening more effective, because it brings the holes in the substrate closer to the charge they are screening.  As a result, both $\Delta_{Au}$ and $\Delta_{aSi}$ decrease as $V_{aSi}$ is made more negative \cite{Inversion}.

The response of $G_{M}$ to the gold $\Delta_{Au}$ decreases as $V_{aSi}$ is made more negative until $V_{aSi} \approx$ -8 V, at which point it saturates.  This saturation is expected, because once the depletion layer below the gold shrinks to zero, so that the Si-SiO$_{2}$ interface underneath the gold is in accumulation, the distance between the charge on the gold and the screening charge is fixed at the SiO$_{2}$ thickness (100 nm). $\Delta_{aSi}$ does not appear to saturate as $V_{aSi}$ is made more negative.  This is not surprising, because the a-Si:H is very close to the MOSFET gate.  Because there must always be a depletion layer between the inversion layer of the MOSFET and the $p$-type substrate, the Si-SiO$_{2}$ interface underneath the a-Si:H cannot be brought into accumulation, and the signal does not saturate.
It is however surprising that for $V_{aSi} <$ -10 V, $\Delta_{Au}$ is larger than $\Delta_{aSi}$.  Although the gold contacts are physically much larger than the a-Si:H strip, which enhances their effect on $G_{M}$ relative to the a-Si:H, the a-Si:H strip is much closer to the MOSFET, so one would not expect $\Delta_{Au}$ ever to be significantly larger than $\Delta_{aSi}$.  Thus, although the dependencies of $\Delta_{aSi}$ and $\Delta_{Au}$ on $V_{aSi}$ can be understood in terms of screening, the relative magnitudes of these quantities are not currently understood.  We have also measured the dependence of $\Delta_{Au}$ and $\Delta_{aSi}$ on $V_{aSi}$ at T = 98 K and T = 179 K.  The results are qualitatively similar, but the relative magnitudes of $\Delta_{aSi}$ and $\Delta_{Au}$ change somewhat depending on the temperature, a result that is also currently not understood.

\begin{figure}
    \setlength{\unitlength}{1cm}
    \begin{center}
        \includegraphics[width=5.5cm, keepaspectratio=true]{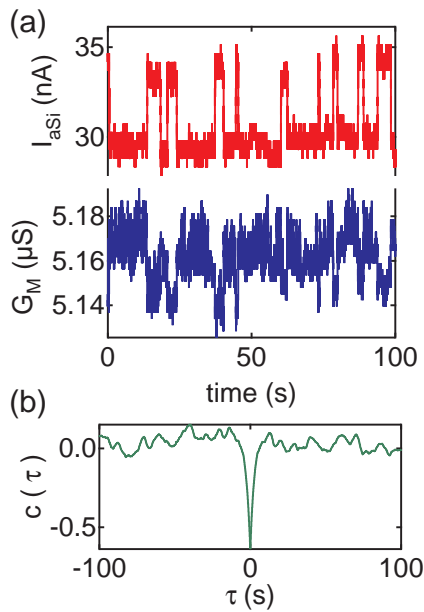}
    \end{center}
    \caption{Noise correlations measured at room temperature. (a) Current through a-Si:H strip $I_{aSi}$ (top trace) and transistor conductance $G_{M}$ (bottom trace) as a function of time. Here we apply a constant voltage
bias of 2 V across the a-Si:H strip. (c) Correlation between $I_{aSi}$ and $G_{M}$, as discussed in the main text.}
    \label{fig:fig4}
\end{figure}	

We have thus seen that screening by holes in the $p$-type substrate decreases the sensitivity of our MOSFET electrometer.  We expect that there are other sources of screening in our system, for instance by the metallic gate of the MOSFET.  Despite the effect of screening, our electrometer is still sensitive to very small charge fluctuations in the a-Si:H, even at room temperature. An intriguing demonstration of this is the sensitivity of the MOSFET to telegraph noise switches in the a-Si:H. $1/f$ noise and discrete telegraph switches have been observed previously in the resistance of macroscopic a-Si:H samples \cite{SwitchaSi}.  The discrete switching that is sometimes observed occurs for samples where the conductance is dominated by filaments small enough to be affected by a single switch.  While the microscopic origin of $1/f$ noise in a-Si:H is unclear, its phenomenology is quite rich, and it is closely connected with Staebler-Wronski effect \cite{aSiH}, as demonstrated in Parman \textit{et al.} \cite{KakaliosNoisePower}.

 At room temperature, where the resistance of the a-Si:H is not too large, we apply a voltage between the two gold a-Si:H contacts and measure the current $I_{aSi}$ that flows through the a-Si:H strip.
The top trace of Fig. \ref{fig:fig4}(a) shows $I_{aSi}$ measured as a function of time, exhibiting clear telegraph noise.  This switching appeared and disappeared apparently randomly, lasting $\sim$ 1 day. Because our sample is nanopatterned, it is not clear whether the origin of the telegraph noise we observe is the same as the origin of the noise found in bulk a-Si:H samples.  However, the conductance of our heavily doped a-Si:H strip is only weakly dependent on the voltages of nearby gates, such as the voltage applied to the MOSFET gate or $p$-type substrate. For example, we find that we must change the MOSFET gate voltage by $\sim$ 30 V in order to produce a change in $I_{aSi}$ as large as the $\sim$ 5 pA fluctuations shown in Fig. \ref{fig:fig4}(a). The narrow a-Si:H strip is thus not very sensitive to its electrostatic environment, and it is therefore likely that the switching seen in Fig. \ref{fig:fig4}(a) results from fluctuations inside or on the surface of the a-Si:H, as opposed to electron trapping external to the a-Si:H.

As we measure $I_{aSi}$(t), we simultaneously measure $G_{M}$(t), and the results are plotted in the bottom trace of  Fig. \ref{fig:fig4}(a). We see that $I_{aSi}$ and $G_{M}$ are anti-correlated.  When $I_{aSi}$ jumps up, $G_{M}$ jumps down, and vice versa.  This anti-correlation is demonstrated quantitatively in Fig. \ref{fig:fig4}(b).  Here we measure $I_{aSi}$ and $G_{M}$ simultaneously for a much longer time than shown in Fig. \ref{fig:fig4}(a), and compute the cross-correlation function between the two signals $c(\tau)$ \cite{RandomSignal}.  Here we have normalized $c(\tau)$  by subtracting the product of the means of $I_{aSi}$ and $G_{M}$, and then dividing by the product of their standard deviations \cite{CorrNorm}.  We see that for our data $c(\tau)$ has a negative peak at $\tau$ = 0 with a value $\approx -0.6$, indicating that the two signals $I_{aSi}$ and $G_{M}$ are highly anti-correlated: With our normalization $c(0) = -1$ corresponds to perfect anti-correlation.

From these data, it is clear that the MOSFET electrometer can detect single switches in a material adjacent to it.  It may be that electrostatic fluctuations that give rise to the switching noise in the a-Si:H current are detected by the MOSFET directly, or that these fluctuations change the charge distribution along the a-Si:H strip to which the MOSFET is extremely sensitive.  We have observed telegraph noise in the current through nanopatterned strips of a-Si:H other than the one studied here, but these samples were not fabricated adjacent to a MOSFET charge sensor.  The intermittency of the switch investigated here made it difficult to study in detail, and more work is required to determine the mechanism by which the MOSFET senses these switches.

In summary, we have shown experimentally that electrostatic screening significantly affects the charge sensitivity of a nanometer scale electrometer and that despite this effect, the electrometer is still very sensitive to its electrostatic environment, even at room temperature.  We expect that this work will be used to help mitigate the effects of screening in the development of even more sensitive nanoscale electrometers.

	This work was supported by the Department of Energy under Award Number DE-FG02-08ER46515 and in part by the US Army Research Office under Contract W911NF-07-D-0004.


\begin{thebibliography}{21}
\bibitem[{sub()}]{substrate}
\bibinfo{note}{We specify the voltages applied to the gold contacts relative to
  the voltage applied to the $p$-type substrate.}
\bibitem[{Inv()}]{Inversion}
\bibinfo{note}{Here we observe a monotonic decrease in $\Delta_{aSi}$ with
  increasingly negative $V_{aSi}$ for $V_{aSi} < $ 0 V. However, if $V_{aSi}$
  is made sufficiently more positive, an inversion layer will form underneath
  the a-Si:H, which we expect may lead to additional screening and a decrease
  in $\Delta_{aSi}$. We have observed such a decrease in $\Delta_{aSi}$ in an
  experiment performed with $V_{sub}$ = 0 V.}
\end{thebibliography}

\end{document}